\documentclass[11pt,english]{amsart}
\usepackage{cite}
\usepackage{hyperref}
\usepackage{url}
\usepackage{babel,enumerate,mathrsfs}
\usepackage{psfrag,graphicx,float}
\usepackage{amsmath,amssymb,amsthm}
\usepackage[mathscr]{eucal}

\def\be{\begin{equation}}
\def\ee{\end{equation}}
\def\ba{\begin{eqnarray}}
\def\ea{\end{eqnarray}}
\def\bs{\begin{subequations}}
\def\es{\end{subequations}}

\newfont{\bbd}{msbm10 scaled\magstep1}

\def\beq{\begin{equation}}
\def\eeq{\end{equation}}
\def\bea{\begin{eqnarray}}
\def\eea{\end{eqnarray}}
\newtheorem{definition}{Definition}[section]

\def\ep{\varepsilon}
\def\ri{{\rm{i}}}

\newcommand{\CA}{{\mathcal A}}
\newcommand{\CM}{{\mathcal M}}

\def\be{\beta}

\def\ep{\varepsilon}

\newcommand{\rref}[1]{(\ref{#1})} 

\def\mat2#1#2#3#4{{\left(\begin{array}{cc}#1 & #2\\ #3 & #4
      \end{array}\right)}}

\def\mats2#1#2#3#4{{\left(\begin{array}{cc}#1 & #2\vspace{2truemm} \\ #3 & #4
\end{array}\right)}}

\def\tilde{\widetilde}

\def\ri{{\rm{i}}}

\def\su2{{\mathfrak{su}(2)}}

\begin{document}
\title[Integrability of differential--difference equations ]{Integrability of  differential--difference equations  with discrete kinks}
\author{Christian Scimiterna}
\email{Scimiterna@fis.uniroma3.it}
\address{Dipartimento di Ingegneria Elettronica, Universit\`{a} di Roma Tre, and INFN, Sezione di
Roma Tre, via della Vasca Navale, 84, Roma, Italy}
\author{Decio Levi}
\email{levi@roma3.infn.it}
\address{Dipartimento di Ingegneria Elettronica, Universit\`{a} di Roma Tre, and INFN, Sezione di
Roma Tre, via della Vasca Navale, 84, Roma, Italy}

\begin{abstract}
In this article we discuss a series of  models introduced by Barashenkov, Oxtoby and Pelinovsky to describe some discrete approximations to the $\phi^4$ theory which preserve travelling kink solutions. We show, by applying the multiple scale test that they have some integrability properties as they pass the $A_1$ and $A_2$ conditions. However they are not integrable as they fail the $A_3$ conditions.

\end{abstract}
\date{\today}
\maketitle
\section{Introduction}

In a recent article Barashenkov, Oxtoby and Pelinovsky \cite{bop} considered  those exceptional discretizations of the $\phi^4$ theory in which the stationary kink can be
centered anywhere between the lattice sites.  The request that  the kink be translational invariance 
generates three families of exceptional discretizations which provide new differential--difference equations which   are not known to be integrable. The continuous $\phi^4$ model in itself is not completely integrable, i.e. it has not an infinite number of generalized symmetries, conservation laws and exact solutions, however it is an approximation of many $C$ or $S$ integrable nonlinear models like the Liouville equation or the sine--Gordon equation.  So the analysis of their integrability is an interesting problem worthwhile to be investigated. To do so we apply the multiple scale analysis to the nonlinear discrete equations as this provides an integrability test and gives an integrability grading of the equations.

In Sections   and \ref{sem} and \ref{msele} we present a review of the models presented in \cite{bop} and of the results necessary to apply the multiple scale analysis to them. Then in Section \ref{sme} we present the results of the calculations and in the last Section we give some concluding remarks. The details of the calculations are left to an Appendix.  

\section{Exceptional models} \label{sem}
$\phi^4$ theory has been  one of the most important nonlinear models for the description of statistical mechanics and field theory systems \cite{bs,r}. The discrete analogs of the $\phi^4$ kinks are solutions of  equations of the form
\bea \label{a1}
\ddot u_n = \frac{u_{n+1} - 2 u_n + u_{n-1}}{h^2} + \frac{u_n}{2}+Q_n( u_{n-1}, u_n, u_{n+1}),
\eea 
where $u_n=u(hn,t)$, with $h$ a constant lattice spacing.  The function $Q_n$ is chosen so as to  give in the continuous limit, at first order in $h$, the  $\phi^4$ potential $Q=-\frac{1}{2} u^3$. These kinks
have been used to
describe charge-density waves in polymers and metals \cite{4}, narrow domain walls in ferroelectrics \cite{5}, discommensurations in dielectric crystals \cite{6}, and topological excitations in hydrogen-bonded chains \cite{7,8}.
Physically, one of the most significant properties of domain
walls and topological defects is their mobility \cite{12}. 
Mathematically,
 discrete equations \rref{a1} are known to admit
			stationary kink solutions \cite{12,13}; however whether traveling
discrete kinks exist remains  an open question \cite{3,14,15,16,17,18}. The discretization breaks the Lorentz
invariance and the existence of traveling discrete kinks
becomes a nontrivial matter. Indeed, if the equation is non authonomous, the discretization even breaks the translation invariance
of  \rref{a1} due to the
presence of a Peierls-Nabarro barrier, an additional periodic
potential induced by discreteness.  Miraculously, there are several exceptional discretizations
which, while breaking the translation invariance of the equation,
allow  for the existence of translationally invariant kinks,
that is, kinks centered at an arbitrary point between the sites.
One such discretization was discovered by Speight and Ward
	using a Bogomolny-type energy-minimality argument \cite{15,16}.
In \cite{bop} we have a systematic study of these exceptional cases based on the observation  that the translational invariance of the kink implies the
existence of an underlying one-dimensional map $u_{n+1}=F(u_n)$. A simple algorithm based on this observation gives three classes of  exceptional discretizations:
\begin{subequations} \label{a2}
\bea 
Q_{1}&\doteq&\frac{1} {20} [\mu_{1} (u_{n+1}+u_{n-1} ) (u_{n+1}^{2}+u_{n-1}^{2} )+ \\
\nonumber &+&\mu_{3}u_{n}^{2} (u_{n+1}+u_{n-1} )+\mu_{2}u_{n}(u_{n+1}^{2}+u_{n}^{2}+u_{n-1}^{2}+u_{n+1}u_{n-1} ) ], \\
Q_{2}&\doteq&\mu_{1}^{2} (u_{n+1}^{3}+u_{n-1}^{3} )+2\mu_{3} (\mu_{1}-\mu_{2} )u_{n+1}u_{n}u_{n-1}+ \\
\nonumber&+&\mu_{1}\left(\mu_{1}-\mu_{2}\right)u_{n+1}u_{n-1}\left(u_{n+1}+u_{n-1}\right)+\mu_{1}\left(2\mu_{3}+\mu_{2}\right)u_{n}\left(u_{n+1}^{2}+u_{n-1}^{2}\right)+\\
\nonumber&+&2\mu_{1}\left(\mu_{2}+\mu_{3}\right)u_{n}^{3}+\left[2\mu_{1}^{2}+\mu_{3}^{2}+\mu_{2}\left(\mu_{3}-\mu_{1}\right)\right]u_{n}^{2}\left(u_{n+1}+u_{n-1}\right),\\
 Q_{3}&\doteq&\mu_{1}\left(1+\frac{h^2} {4}\right)u_{n}\left(u_{n+1}+u_{n-1}\right)^{2}-2\mu_{1}u_{n-1}u_{n+1}\left(u_{n+1}+u_{n-1}\right)+\\
\nonumber&+&\left(\frac{1} {4}-\frac{\mu_{2}} {2}\right)u_{n}^{2}\left(u_{n+1}+u_{n-1}\right)+\mu_{2}u_{n-1}u_{n}u_{n+1}. 
\eea
\end{subequations}
where in the first two equations  the coefficients ($\mu_1, \mu_2, \mu_3$), as presented in \cite{bop},  are not independent. 
\section{Multiple scale expansion of lattice equations} \label{msele}
For completeness we present here the basic ideas of the multiple scale expansion of lattice equations as presented in \cite{ls_dNLS} in the case of expansions in terms of analytic functions.
\subsection{Expansion of real dispersive partial difference equations}

\subsubsection{From shifts to derivatives}

We consider a function $u_{n}: \mathbb{Z}\rightarrow \mathbb{R}$ depending on a discrete index $n\in\mathbb{Z}$ and  suppose that: 
\begin{enumerate} [(a)]

\item The dependence of $u_{n}$ on $n$ is realized through the \emph{slow variable} $n_{1}\doteq\ep n\in\mathbb{R}$, $\ep\in\mathbb{R}$, $0<\ep\ll 1$, i.e. $u_{n}\doteq u(n_{1})$;

\item  $n_{1}$ varies in a region of the integer axis such that $u\left(n_{1}\right)$ is therein analytical;

\item The radius of convergence of the Taylor series  in $ n_{1}$ is wide enough to include as  \emph{inner points} all the points involved in the discrete equation.

\end{enumerate}
Under these hypotheses we can write the action of the shift operator $T_{n}$, defined by  $T_{n} u_{n}\doteq u_{n+1}=u(n_{1}+\ep)$ in  the variable $ n_{1}$, as
\begin{eqnarray} \label{e1}
T_{n} u( n_{1})&=&u( n_{1})+\ep u^{(1)}( n_{1})+\frac{\ep^2}{2
}u^{(2)}( n_{1})+ ... +\frac{\ep^i}{i!}u^{(i)}( n_{1})+ ...\\ \nonumber &=&\sum_{i=0}^{+\infty}\frac{\ep^i}{i!} u^{(i)}( n_{1}),
\end{eqnarray}
where $u^{(i)}( n_{1})\doteq \frac{d^{i} u(n_{1})}{ dn_{1}^{i}}\doteq d_{n_{1}}^{i} u( n_{1})$, being $d_{n_{1}}$ the total derivative operator. From  \rref{e1} it follows that we can write  the following formal expansion for the shift operator $T_n$:
\begin{eqnarray}
T_{n}=\sum_{i=0}^{+\infty}\frac{\ep^i}{i!}\,d_{n_{1}}^i\doteq e^{\ep d_{n_{1}}}.\label{Larth}
\end{eqnarray}
If $u_n$  depends simultaneously on the \emph{fast variable} $n$ and on the \emph{slow variable} $n_{1}$ i.e. $u_{n}\doteq u(n,n_{1})$, the action of the \emph{total} shift operator $T_{n}$ will  give $T_{n} u_{n}\doteq u_{n+1}=u(n+1,n_{1}+\ep)$. So we can split it in terms of \emph{partial} shift operators $\mathcal{T}_{n}$ and $\mathcal{T}_{n_{1}}^{(\ep)}$ which are defined respectively by $\mathcal{T}_{n}u(n,n_{1})=u(n+1,n_{1})$ and $\mathcal{T}_{n_{1}}^{(\ep)}u(n,n_{1})=u(n,n_{1}+\ep)$, 
$
T_{n}\doteq\mathcal{T}_{n}\mathcal{T}_{n_{1}}^{(\ep)},
$
 where $\mathcal{T}_{n_{1}}^{(\ep)}$ is also given by Eq. (\ref{Larth}). The dependence of $u_{n}$ on $n$ can be easily extended to the case of one fast variable $n$ and $K$ slow variables $n_{j}\doteq\ep_{j}n$, $\ep_{j}\in\mathbb{R}$, $1\leq j\leq K$. The  total shift operator $T_n$ will now be written as:
$
T_{n}\doteq\mathcal{T}_{n}\prod_{j=1}^K\mathcal{T}_{n_{j}}^{(\ep_{j})}.
$

 Let us now  consider a nonlinear partial difference equation
\begin{eqnarray}
F\left[u_{\left\{n+i\right\}_{i=-{\mathcal N}^{(-)}}^{{\mathcal N}^{(+)}},\left\{m+j\right\}_{j=-\CM^{(-)}}^{\CM^{(+)}}}\right]=0,\ \ \ \ \ {\mathcal N}^{(\pm)},\ \CM^{(\pm)}\geq 0,\label{Esperia}
\end{eqnarray}
for a function $u_{n,m}: \mathbb{Z}^2\rightarrow \mathbb{R}$ which depends on two integer indexes $n$ and $m$ which  we will call respectively \emph{space} and \emph{time} variables. Eq. (\ref{Esperia}) contains shifts of $m$  contained in the intervals  ($m-\CM^{(-)}$, $m+\CM^{(+)}$) and and $n$-shifts in the interval  ($n-{\mathcal N}^{(-)}$, $n+{\mathcal N}^{(+)}$). 
 Under the hypotheses (a, b, c)  we  can give a series representation of the shifted values of $u_{n,m}$ around the point $\left( n, m\right)$. Choosing
\begin{eqnarray}
\nonumber\ep_{n_{1}}\doteq N_{1}\ep,\ \ \ \ \ \ep_{m_{j}}\doteq M_{j}\ep^{j},\ \ \ \ \ 1\leq j\leq K,\ \ \ \ \ \ (\ep,\, N_1, \, M_j) \in R
\end{eqnarray}
 we can write
\begin{subequations}\label{Istria}
\begin{eqnarray}
&&T_{n}=\mathcal{T}_{n}\mathcal{T}_{n_{1}}^{(\ep_{n_{1}})}=\mathcal{T}_{n}\sum_{j=0}^{+\infty}\ep^j\CA_{n}^{(j)},\ \ \ \ \ \CA_{n}^{(j)}\doteq\frac{N_{1}^{j}} {j!}\partial_{n_{1}}^{j},\label{Istria1}\\
&&T_{m}=\mathcal{T}_{m}\prod_{j=1}^K\mathcal{T}_{m_{j}}^{(\ep_{m_{j}})}=
\mathcal{T}_{m}\sum_{j=0}^{+\infty}\ep^j\CA_{m}^{(j)}, \\
&&T_{n}T_{m}=\mathcal{T}_{n}\mathcal{T}_{m}\mathcal{T}_{n_{1}}^{(\ep_{n_{1}})}\prod_{j=1}^K\mathcal{T}_{m_{j}}^{(\ep_{m_{j}})}=\mathcal{T}_n\mathcal{T}_m\sum_{j=0}^{+\infty}\ep^j\CA_{n,m}^{(j)},
\end{eqnarray}
\end{subequations}
 where the operators  $\CA_{m}^{(j)}$, $\CA_{n,m}^{(j)}$ are appropriate combinations of $ \CA_{m_k}^{(j)}\doteq\frac{M_{k}^{j}}{j!}\partial_{m_{k}}^{j}$ and $\CA_{n}^{(j)}\doteq\frac{N_{1}^{j}} {j!}\partial_{n_{1}}^{j}$  (see \cite{tS}~ for more details). When we insert  the explicit expressions (\ref{Istria}) of the shift operators in term of the derivatives with respect to the slow variables into eq. (\ref{Esperia}) we get an $\epsilon$--dependent  PDE of infinite order. 
Then we assume for the function $u=u(n,m,n_{1},\left\{m_{j}\right\}_{j=1}^{K},\ep)$ a double expansion in harmonics and in the perturbative parameter $\ep$
\begin{eqnarray}
u\left(n,m,n_{1},\left\{m_{j}\right\}_{j=1}^{K},\ep\right)=\sum_{\gamma=1}^{+\infty}\sum_{\alpha=-\gamma}^{\gamma}\ep^{\gamma}u^{(\alpha)}_{\gamma}\left(n_{1},\left\{m_{j}\right\}_{j=1}^K\right)E_{n,m}^{\alpha},\label{Ilio}\\
\nonumber E_{n,m}\doteq e^{i\left[\kappa n-\omega\left(\kappa\right) m\right]},\ \ \ \ \ u^{(-\alpha)}_{\gamma}=\bar u^{(\alpha)}_{\gamma}\ \ \ \ \ \ \ \ \ \ \ \ \ \ \ \ \ 
\end{eqnarray}
where the index $\gamma$ is chosen $\geq 1$ in order to let any nonlinear part of eq. (\ref{Esperia}) to enter as a perturbation in the multiscale expansion.

\subsubsection{From derivatives to shifts}

Splitting (\ref{Esperia}) in the various powers of $\epsilon$ and in the different harmonics the multiple scale approach produces from a given partial difference equation  partial differential equations for the amplitudes $u^{(\alpha)}_{\gamma}$. Starting from the obtained partial differential equation we can write down a partial difference equation inverting the expression of the shift operator as
\begin{eqnarray}\label{i1}
\partial_{n_{1}}=ln\mathcal{T}_{n_{1}}=ln\left(1+h_{n_1}\Delta^{(+)}_{n_{1}}\right)\doteq\sum_{i=1}^{+\infty}\frac{(-1)^{i-1}h_{n_1}^i} {i}\Delta^{(+)i}_{n_{1}},
\end{eqnarray}
where $\Delta^{(+)}_{n_{1}}\doteq(\mathcal{T}_{n_{1}}-1)/h_{n_1}$ is just the first \emph{forward} difference operator with respect to the slow-variable $n_{1}$. 
 
 Only when we impose that the function $u_n$ is a slow--varying function of order $l$ in $n_1$, i.e. $\Delta_{n_1}^{\ell+1} u_n \approx 0$,  the $\partial_{n_{1}}$ operator  reduces to polynomials in  $\Delta_{n_{1}}$ of order at most $l$. In \cite{levi}, choosing $l=2$ for the indexes $n_{1}$ and $m_{1}$ and $l=1$ for $m_{2}$, it was shown that the integrable \emph{lattice potential $KdV$} equation \cite{NC} reduces to a completely discrete and local Nonlinear Schr\"odinger Equation ($NLSE$) which has been proved to be not integrable by singularity confinement and algebraic entropy \cite{viallet,ramani}. Consequently, if one  passes from derivatives to shifts, one ends up in general with a \emph{nonlocal} partial difference equation in the slow variables $n_{\kappa}$ and $m_{\delta}$.

\subsection{The orders beyond the $NLSE$ equation and the integrability conditions}

The multiscale expansion of a difference equation  (\ref{Esperia}) for analytic functions will give rise to a continuos PDE. So a multiple scale integrability test will require that an equation of  the class of equations  (\ref{a1}) is integrable if its multiscale expansion will go into the hierarchy of the $NLSE$. To be able to do so we need to consider here  the orders beyond that at which one obtains for the harmonic $u^{\left(1\right)}_{1}$ the (integrable) $NLSE$.   The first attempts to go beyond the $NLSE$ order has been presented by Kodama and Mikhailov and by Santini, Degasperis and Manakov in \cite{MK,DMS}. Starting from $S$ integrable models (models integrable via a Scattering Transform), through a combination of an asymptotic functional analysis and spectral methods, one succeeds in removing all the secular terms from the reduced equations order by order. The results could be summarized as follows:

\begin{enumerate}

\item The number of slow-time variables required to $\cdots$ for the amplitudes $u^{\left(\alpha\right)}_{j}$s  coincides with the number of nonvanishing coefficients $\omega_{j}\left(\kappa\right)=\frac{1}{j!}\frac{d^j \omega(k)}{dk^j}$;

\item The amplitude $u^{\left(1\right)}_{1}$ evolves at the slow-times $t_{\sigma}$, $\sigma\geq 3$ according to the $\sigma-$th equation of the $NLS$ hierarchy;

\item The amplitudes of the higher perturbations of the first harmonic $u^{\left(1\right)}_{j}$, $j\geq 2$,  taking into account some \emph{asymptotic boundary conditions}, evolve at the slow-times $t_{\sigma}$, $\sigma\geq 2$ according to certain \emph{linear, nonhomogeneous} equations.

\end{enumerate}

\noindent Thus  the cancellation at each stage of the perturbation process of all the secular terms  is a sufficient condition  to uniquely fix the evolution equations followed by every $u^{\left(1\right)}_{j}$, $j\geq 1$ for each slow-time $t_\sigma$. Point 2 implies that a hierarchy of integrable equations always provide compatible evolutions for a unique function $u$ depending on different times, i.e. the equations in its hierarchy are \emph{generalized symmetries} of each other. 
In this way this procedure provides  \emph{necessary and sufficient} conditions to get secularity-free reduced equations \cite{DMS}.

Then, following Degasperis and Procesi \cite{DP} we state:
\begin{definition}
A nonlinear PDE is said to be integrable if it possesses a nontrivial Lax pair and consequently an infinity of generalized symmetries.
\end{definition}
As a consequence of this Definition we have the following Proposition:

\noindent {\bf Proposition 3.1} {\it 
If equation (\ref{Esperia}) is $S$ integrable, then under a multiscale expansion the functions $u^{\left(1\right)}_{j}$, $j\geq1$ satisfy the equations}
\begin{subequations}\label{Valentia}
\begin{eqnarray}
\partial_{t_{\sigma}}u^{\left(1\right)}_{1}=K_{\sigma}\left[u^{\left(1\right)}_{1}\right],\label{Valentia1}\ \ \ \ \ \ \ \ \ \ \ \ \ \\
M_{\sigma}u^{\left(1\right)}_{j}=f_{\sigma}(j),\ \ \ M_{\sigma}\doteq\partial_{t_{\sigma}}-K_{\sigma}^{\prime}\left[u^{\left(1\right)}_{1}\right],\label{Valentia2}
\end{eqnarray}
\end{subequations}
 \emph{$\forall\ j,\ \sigma\geq 2$, where $K_{\sigma}\left[u^{\left(1\right)}_{1}\right]$ is the $\sigma$--th flow in the nonlinear Schr\" odinger hierarchy. All the other  $u_{j}^{(\kappa)}$, $\kappa\geq 2$ are expressed in terms of differential monomials of $u_{\rho}^{(1)}$, $\rho\leq j$.}\\

\noindent In (\ref{Valentia2})  $f_{\sigma}(j)$ is a nonhomogeneous \emph{nonlinear} forcing term and $K_{\sigma}^{\prime}\left[u\right]v$ is the Frechet derivative of the nonlinear term $K_{\sigma}[u]$ along the direction $v$ $K_{\sigma}^{\prime}[u]v\doteq\frac{d} {ds}K_{\sigma}[u+sv]\mid_{s=0}$,
 i. e. the linearization  of $K_{\sigma}[u]$ along the direction $v$. Eqs. (\ref{Valentia}) are a \emph{necessary} condition for integrability and represent a hierarchy of \emph{compatible} evolutions for the same function $u^{\left(1\right)}_{1}$ at different slow-times.  The compatibility of eqs. (\ref{Valentia2}) is not always guaranteed but is subject to a sort of commutativity conditions among their r. h. s. terms $f_{\sigma}(j)$. Then it is easy to prove that the operators $M_{\sigma}$ defined in eq. (\ref{Valentia2}) commute among themselves. 
 Once we fix the index $j\geq 2$ in the set of eqs. (\ref{Valentia2}), this commutativity condition implies the following \emph{compatibility} conditions
\begin{eqnarray}
M_{\sigma}f_{\sigma'}\left(j\right)=M_{\sigma'}f_{\sigma}\left(j\right),\ \ \ \forall\, \sigma,\sigma'\geq 2,\label{Lavinia}
\end{eqnarray}
where, as $f_{\sigma}\left(j\right)$ and $f_{\sigma'}\left(j\right)$ are functions of the different perturbations of the fundamental harmonic up to degree $j-1$, the time derivatives $\partial_{t_{\sigma}}$, $\partial_{t_{\sigma'}}$ of those harmonics appearing respectively in $M_{\sigma}$ and $M_{\sigma'}$ have to be eliminated using the evolution equations (\ref{Valentia}) up to the index $j-1$. \underline{ The commutativity conditions \rref{Lavinia} turn out to be an integrability test.} 

Following \cite{DMS} we conjecture
 that the relations (\ref{Valentia}) are a \emph{sufficient} condition for the integrability or that the integrability is a \emph{necessary} condition to have a multiscale expansion where eqs. (\ref{Valentia}) are satisfied. To construct the functions $f_{\sigma}(j)$ according to the Proposition 3.1 we define:
\begin{definition}
{\it A differential monomial $\rho\left[u^{\left(1\right)}_{j}\right]$, $j\geq 1$ in the functions $u^{\left(1\right)}_{j}$, its complex conjugate and its $\xi$-derivatives is a monomial of "gauge" 1 if it possesses the transformation property
\begin{eqnarray}
\nonumber\rho\left[\tilde u^{\left(1\right)}_{j}\right]= e^{\ri\theta}\rho\left[u^{\left(1\right)}_{j}\right],\ \ \ \ \ \tilde u^{\left(1\right)}_{j}\doteq e^{\ri\theta}u^{\left(1\right)}_{j};
\end{eqnarray}}
\end{definition}
\begin{definition}\label{FAVL1}
{\it A finite dimensional vector space $\mathcal P_{n}$, $n\geq 2$ is the set of all differential polynomials in the functions $u^{\left(1\right)}_{j}$, $j\geq 1$, their complex conjugates and their $\xi$-derivatives of order $n$ in $\ep$ and gauge 1 when 
\begin{eqnarray}
\mbox{order}\left(\partial_{\xi}^{m}u^{\left(1\right)}_{j}\right)=\mbox{order}\left(\partial_{\xi}^{m}\bar u^{\left(1\right)}_{j}\right)=m+j,\ \ \ m\geq 0;\nonumber
\end{eqnarray}}
\end{definition}
\begin{definition}\label{FAVL2}
{\it $\mathcal P_{n}(m)$, $m\geq 1$ and $n\geq 2$ is the subspace of $\mathcal P_{n}$ whose elements are differential polynomials in the functions $u^{\left(1\right)}_{j}$s, their complex conjugates and their $\xi$-derivatives of order $n$ in $\ep$ and gauge 1 for $1\leq j\leq m$.}
\end{definition}
From the definition (\ref{FAVL2})  one can see that in general $K_{\sigma}\left[u^{\left(1\right)}_{1}\right]\in\partial_{\xi}^\sigma u^{\left(1\right)}_{1}\cup\mathcal P_{\sigma+1}(1)$ and that $f_{\sigma}(j)\in\mathcal P_{j+\sigma}(j-1)$ where $j,\ \sigma\geq 2$. The basis monomials of the spaces $\mathcal P_{n}(m)$ can be found, for example, in \cite{tS} 

\noindent {\bf Proposition 3.2} {\it 
If for each fixed $j\geq 2$ the equation (\ref{Lavinia}) with $\sigma=2$ and $\sigma'=3$, namely $M_{2}f_{3}\left(j\right)=M_{3}f_{2}\left(j\right)$, is satisfied, then there exist unique differential polynomials $f_{\sigma}(j)$ $\forall\, \sigma\geq 4$ such that the flows $M_{\sigma}u^{\left(1\right)}_{j}=f_{\sigma}\left(j\right)$ commute for any $\sigma\geq 2$.}

\noindent Hence among the relations (\ref{Lavinia}) only those with $\sigma=2$ and $\sigma'=2$ have to be tested.

\noindent{\bf Proposition 3.3} {\it 
The homogeneous equation $M_{\sigma}u=0$ has no solution $u$ in the vector space $\mathcal P_{m}$, i.e. $Ker\left(M_{\sigma}\right)\cap\mathcal P_{m}=\emptyset$.}

\noindent Consequently the multiscale expansion (\ref{Valentia}) is \emph{secularity-free}. Finally we define the \emph{degree of integrability} of a given equation:
\begin{definition}\label{FrancescoColonnaRomano}
{\it If the relations (\ref{Lavinia}) are satisfied up to the index $j$, $j\geq 2$, we say that our equation is asymptotically integrable of degree $j$ or $A_{j}$ integrable.}
\end{definition}

\subsubsection{Integrability conditions for the $NLSE$ hierarchy}

We specify here  the conditions for asymptotic integrability of order $k$ or $A_{k}$ integrability conditions. To simplify the notation, we will use for $u^{\left(1\right)}_{j}$ the concise form $u(j)$. Moreover, for  convenience of the reader, we list the fluxes $K_{\sigma}\left[u\right]$ of the $NLSE$ hierarchy for $u$ up to $\sigma=4$:

\begin{subequations}\label{Rutuli}
\begin{eqnarray}
K_{1}[u]\doteq Au_{\xi},\ \ \ \ \ \ \ \ \ \ \ \ \ \ \ \ \ \ \ \ \ \ \ \ \ \ \ \ \ \ \ \ \ \ \ \ \ \ \\
K_{2}[u]\doteq-\ri\rho_{1}\left[u_{\xi\xi}+\frac{\rho_{2}} {\rho_{1}}|u|^2u \right],\ \ \ \ \ \ \ \ \ \ \ \ \ \ \ \ \ \ \ \ \ \ \ \ \ \ \ \ \label{Rutuli1}\\
K_{3}[u]\doteq B\left[u_{\xi\xi\xi}+\frac{3\rho_{2}} {\rho_{1}}|u|^2u_{\xi}\right],\ \ \ \ \ \ \ \ \ \ \ \ \ \ \ \ \ \ \ \ \ \ \ \ \ \ \ \ \label{Rutuli2}\\
K_{4}[u]\doteq-\ri C\left\{u_{\xi\xi\xi\xi}+\frac{\rho_{2}} {\rho_{1}}\left[\frac{3\rho_{2}} {2\rho_{1}}|u|^4u+4|u|^2u_{\xi\xi}+3u_{\xi}^2\bar u+2|u_{\xi}|^2u+u^2\bar u_{\xi\xi}\right]\right\},\label{Rutuli3}
\end{eqnarray}
\end{subequations}
where $\rho_{1}$, $\rho_{2}$, $A$, $B$ and $C$ are arbitrary complex constants.

 The $A_{1}$ integrability condition is given by the reality of the coefficient $\rho_{2}$ of the nonlinear term in the $NLSE$. 

 The $A_{2}$ integrability conditions  are obtained choosing $j=2$ in the compatibility conditions (\ref{Lavinia}) with $\sigma=2$ and $\sigma'=3$
\begin{eqnarray}
M_{2}f_{3}\left(j\right)=M_{3}f_{2}\left(j\right).\label{Turno}
\end{eqnarray}
In this case we have that $f_{2}(2)\in\mathcal P_{4}(1)$ and $f_{3}(2)\in\mathcal P_{5}(1)$ with ${\rm dim}(\mathcal P_{4}(1))=2$ and ${\rm dim}(\mathcal P_{5}(1))=5$, so that $f_{2}(2)$ and $f_{3}(2)$ will be respectively identified by 2 and 5 complex constants
\begin{subequations}
\begin{eqnarray}
&& f_{2}(2)\doteq au_{\xi}(1)|u(1)|^2+b\bar u_{\xi}(1)u(1)^2,\label{Abruzzo1}\\
&& f_{3}(2)\doteq\alpha |u(1)|^4u(1)+\beta |u_{\xi}(1)|^2u(1)+\gamma u_{\xi}(1)^2\bar u(1)+\delta\bar u_{\xi\xi}(1)u(1)^2+\epsilon |u(1)|^2u_{\xi\xi}(1).\label{Abruzzo2}
\end{eqnarray}
\end{subequations}
In this way, if $\rho_{2}\not=0$, eliminating from eq. (\ref{Turno})  the derivatives of $u(1)$ with respect to the slow-times $t_{2}$ and $t_{3}$ using the evolutions (\ref{Valentia1}) with $\sigma=2$ and $\sigma'=3$ and equating term by term, we obtain the $A_{2}$ integrability conditions
\begin{eqnarray}
a=\bar a,\ \ \ b=\bar b.\label{CieloUrbico}
\end{eqnarray}
 So at this stage we have  two  conditions obtained requiring the reality of the coefficients $a$ and $b$. The expression of $\alpha$, $\beta$, $\alpha$, $\delta$ in terms of $a$ and $b$ are:
\begin{eqnarray}
\alpha=\frac{3\ri B\rho_{2}a} {4\rho_{1}^2},\ \ \ \beta=\frac{3\ri Bb} {\rho_{1}},\ \ \ \gamma=\frac{3\ri Ba} {2\rho_{1}},\ \ \ \delta=0,\ \ \ \epsilon=\gamma.\label{Molise}
\end{eqnarray}
  
 The $A_{3}$ integrability conditions are derived in a similar way setting $j=3$ in eq. (\ref{Turno}). In this case we have that $f_{2}(3)\in\mathcal P_{5}(2)$ and $f_{3}(3)\in\mathcal P_{6}(2)$ with ${\rm dim}(\mathcal P_{5}(2))=12$ and ${\rm dim}(\mathcal P_{6}(2))=26$, so that $f_{2}(3)$ and $f_{3}(3)$ will be respectively identified by 12 and 26 complex constants
\begin{subequations} \label{f23}
\begin{eqnarray} 
&& f_{2}(3)\doteq\tau_{1}|u(1)|^4u(1)+\tau_{2}|u_{\xi}(1)|^2u(1)+\tau_{3}|u(1)|^2u_{\xi\xi}(1)+\tau_{4}\bar u_{\xi\xi}(1)u(1)^2+\tau_{5}u_{\xi}(1)^2\bar u(1)+\nonumber\\
&&\ \ \ \ \ \ \ \ +\tau_{6}u_{\xi}(2)|u(1)|^2+\tau_{7}\bar u_{\xi}(2)u(1)^2+\tau_{8}u(2)^2\bar u(1)+\tau_{9}|u(2)|^2u(1)+\tau_{10}u(2)u_{\xi}(1)\bar u(1)+\nonumber\\
&&\ \ \ \ \ \ \ \ +\tau_{11}u(2)\bar u_{\xi}(1)u(1)+\tau_{12}\bar u(2)u_{\xi}(1)u(1),\label{Lazio4}\\
&& f_{3}(3)\doteq\gamma_{1}|u(1)|^4u_{\xi}(1)+\gamma_{2}|u(1)|^2u(1)^2\bar u_{\xi}(1)+\gamma_{3}|u(1)|^2u_{\xi\xi\xi}(1)+\gamma_{4}u(1)^2\bar u_{\xi\xi\xi}(1)+\nonumber\\
&&\ \ \ \ \ \ \ \ +\gamma_{5}|u_{\xi}(1)|^2u_{\xi}(1)+\gamma_{6}\bar u_{\xi\xi}(1)u_{\xi}(1)u(1)+\gamma_{7}u_{\xi\xi}(1)\bar u_{\xi}(1)u(1)+\gamma_{8}u_{\xi\xi}(1)u_{\xi}(1)\bar u(1)+\nonumber\\
&&\ \ \ \ \ \ \ \ +\gamma_{9}|u(1)|^4u(2)+\gamma_{10}|u(1)|^2u(1)^2\bar u(2)+\gamma_{11}\bar u_{\xi}(1)u(2)^2+\gamma_{12}u_{\xi}(1)|u(2)|^2+\nonumber\\
&&\ \ \ \ \ \ \ \ +\gamma_{13}|u_{\xi}(1)|^2u(2)+\gamma_{14}|u(2)|^2u(2)+\gamma_{15}u_{\xi}(1)^{2}\bar u(2)+\gamma_{16}|u(1)|^2u_{\xi\xi}(2)+\nonumber\\
&&\ \ \ \ \ \ \ \ +\gamma_{17}u(1)^2\bar u_{\xi\xi}(2)+\gamma_{18}u(2)\bar u_{\xi\xi}(1)u(1)+\gamma_{19}u(2)u_{\xi\xi}(1)\bar u(1)+\gamma_{20}\bar u(2)u_{\xi\xi}(1)u(1)+\nonumber\\
&&\ \ \ \ \ \ \ \ +\gamma_{21}u(2)u_{\xi}(2)\bar u(1)+\gamma_{22}\bar u(2)u_{\xi}(2)u(1)+\gamma_{23}u_{\xi}(2)u_{\xi}(1)\bar u(1)+\gamma_{24}u_{\xi}(2)\bar u_{\xi}(1)u(1)+\nonumber\\
&&\ \ \ \ \ \ \ \ +\gamma_{25}\bar u_{\xi}(2)u_{\xi}(1)u(1)+\gamma_{26}\bar u_{\xi}(2)u(2)u(1).\label{Lazio5}
\end{eqnarray}
\end{subequations}
Let us eliminate from eq. (\ref{Turno}) with $j=3$ the derivatives of $u(1)$ with respect to the slow-times $t_{2}$ and $t_{3}$ using the evolutions (\ref{Valentia1}) respectively with $\sigma=2$ and $\sigma'=3$ and the same derivatives of  $u(2)$ using the evolutions (\ref{Valentia2}) with $\sigma=2$ and $\sigma'=3$. Let us  equate the remaining terms term by term, if $\rho_{2}\not=0$,  and indicating with $R_{i}$ and $I_{i}$ the real and imaginary parts of $\tau_{i}$, $i=1,\ldots,12$, we obtain the $A_3$ integrability conditions
\begin{eqnarray} 
&&R_{1}=-\frac{aI_{6}} {4\rho_{1}},\ \ \ R_{3}=\frac{(b-a)I_{6}} {2\rho_{2}}-\frac{aI_{12}} {2\rho_{2}},\ \ \ R_{4}=\frac{R_{2}} {2}+\frac{(a-b)I_{6}} {4\rho_{2}}+\frac{aI_{12}} {4\rho_{2}},\nonumber\\
&&R_{5}=\frac{R_{2}} {2}+\frac{(a-b)I_{6}} {4\rho_{2}}+\frac{(2b-a)I_{12}} {4\rho_{2}},\ \ \ R_{6}=-\frac{aI_{8}} {\rho_{2}},\ \ \ R_{7}=R_{12}+\frac{(a-b)I_{8}} {\rho_{2}},\nonumber\\
&&R_{8}=R_{9}=0,\ \ \ R_{10}=R_{12},\ \ \ R_{11}=R_{12}+\frac{(a-2b)I_{8}} {\rho_{2}},\nonumber\\
&&I_{4}=\frac{(b+a)R_{12}} {4\rho_{2}}+\frac{\rho_{1}I_{1}} {\rho_{2}}+\frac{I_{2}-I_{3}-2I_{5}} {4}+\frac{\left[2b(a-b)+a^2\right]I_{8}} {4\rho_{2}^2},\ \ \ I_{7}=0,\nonumber\\
&&I_{9}=2I_{8},\ \ \ I_{10}=I_{12},\ \ \ I_{11}=I_{6}+I_{12}.\label{Siculi}
\end{eqnarray}
 For completeness we give the expressions of the $\gamma_{j}$, $j=1,\ldots,26$ as functions of the $\tau_{i}$, $i=1,\ldots,12$:
\begin{eqnarray}
&&\gamma_{1}=\frac{3B} {8\rho_{1}^2}\left[-2bR_{12}-8\rho_{1}I_{1}+2(I_{2}-2I_{3}-2I_{5})\rho_{2}+\ri (b-5a)I_{6}+\frac{2a^2I_{8}} {\rho_{2}}-3\ri aI_{12}\right],\nonumber\\
&&\gamma_{2}=-\frac{3Ba} {4\rho_{1}^2}\left[\ri I_{6}+\frac{(a-2b)I_{8}} {\rho_{2}}+\tau_{12}\right],\ \ \ \gamma_{3}=\frac{3\ri B\tau_{3}} {2\rho_{1}},\ \ \ \gamma_{4}=0,\ \ \ \gamma_{5}=\frac{3\ri B\tau_{2}} {2\rho_{1}},\nonumber\\
&&\gamma_{6}=\frac{3\ri B\tau_{4}} {\rho_{1}},\ \ \ \gamma_{7}=\gamma_{5},\ \ \ \gamma_{8}=\gamma_{3}+\frac{3\ri B\tau_{5}} {\rho_{1}},\ \ \ \gamma_{9}=-\frac{3B(\rho_{2}I_{6}+3a\ri I_{8})} {4\rho_{1}^2},\nonumber\\
&&\gamma_{10}=\frac{3\ri B\rho_{2}R_{6}} {2\rho_{1}^2},\ \ \ \gamma_{11}=0,\ \ \ \gamma_{12}=\frac{3\ri B\tau_{9}} {2\rho_{1}},\ \ \ \gamma_{13}=\frac{3\ri B\tau_{11}} {2\rho_{1}},\ \ \ \gamma_{14}=0,\ \ \ \gamma_{15}=\frac{3\ri B\tau_{12}} {2\rho_{1}},\nonumber\\
&&\gamma_{16}=\frac{3\ri B\tau_{6}} {2\rho_{1}},\ \ \ \gamma_{17}=\gamma_{18}=0,\ \ \ \gamma_{19}=\frac{3\ri B\tau_{10}} {2\rho_{1}},\ \ \ \gamma_{20}=\gamma_{15},\ \ \ \gamma_{21}=\frac{3\ri B\tau_{8}} {\rho_{1}},\nonumber\\
&&\gamma_{22}=\gamma_{12},\ \ \ \gamma_{23}=\gamma_{16}+\gamma_{19},\ \ \ \gamma_{24}=\gamma_{13},\ \ \ \gamma_{25}=\frac{3\ri B\tau_{7}} {\rho_{1}},\ \ \ \gamma_{26}=0.\label{gam}
\end{eqnarray}

The conditions in the case of $C$ integrable or linearizable equations are similar to those presented here and can be found in ref. \cite{Burgers}.
\section{Multiple scale Expansions of the exceptional models} \label{sme}

\indent Let us perform the multiscale analysis of the differential-difference models (\ref{a1}) for the real function 
$u_{n}\left(t\right)$ with the  functions $Q_n$ given by \rref{a2}. To do so we expand the variable $u_n$ according to the prescriptions of the previous section and split the obtained system in terms of the different harmonics and various powers of $\epsilon$. The equations at the various orders are presented in the Appendix A, here we just present the final results. 

\noindent{\bf Proposition 4.1} {\it All models \rref{a2} pass the $A_1$ and $A_2$ integrability conditions however fail the $A_3$ one for any choice of the involved parameters $\mu_i$.
.}

Moreover, even the particular choice done by Barashenkov, Oxtoby and Pelinovsky  in the case of $Q_1$  and $Q_2$ is no more integrable than the others. 

In Appendix A it is also shown that these nonlinear model can never be nontrivially linearized.

\section{Concluding remarks} \label{concl}
In this article we have shown, applying the multiple scale expansion integrability test, that the set of discrete $\phi^4$ models constructed by Barashenkov, Oxtoby and Pelinovsky  are not integrable but have just  some nontrivial degree of integrability as they pass the two lowest integrability conditions. 

The models do not satisfy the symmetries of the continuous $\phi^4$ theory but just possess travelling kink solutions. It would be interesting to construct discrete  models which preserve the symmetries of the continuous one. This can be done \cite{lw}, however this implies that we will need to have a variable non uniform lattice. As the continuous model is not completely integrable, we do not expect such discrete  symmetry preserving model  to be completely integrable.

An interesting  extension of the calculation done here consists in expressing the  multiple scale expansion in term of normal forms.  In such a way we would be able to use the results to give approximations to the solutions of the starting equations. Work on both these aspect is in progress.

\section*{Acknowledgments}
This research  has been partly supported by the Italian Ministry of Education and Research, PRIN
 ``Continuous and discrete nonlinear integrable evolutions: from water
waves to symplectic maps".
\appendix
\section{Details of the calculations}
We present here in the following the equations one obtains by splitting \rref{a1}, once we substitute the field $u_n$ by its multiple scale expansion \rref{Ilio}, into the various harmonics and different orders of $\ep$. Except when explicitly stated, the equations are valid for any function $Q_n$ given in \rref{a2}.
\begin{itemize}

\item\emph{Order $\ep$ and $\alpha=0$}: One obtains

\begin{eqnarray}
u_{1}^{(0)}=0;\label{Pyrotechnia}
\end{eqnarray}

\item\emph{Order $\ep$ and $\alpha=1$}: If one requires that $u_{1}^{(1)}\not=0$, one obtains the dispersion relation

\begin{eqnarray}
\omega^2=\frac{4\sin^2\left(\kappa h/2\right)} {h^2}-\frac{1} {2}.\label{Armentano}
\end{eqnarray}

\item\emph{Order $\ep^2$ and $\alpha=0$}: We obtain

\begin{eqnarray}
u_{2}^{(0)}=0;\label{Platone}
\end{eqnarray}

\item\emph{Order $\ep^2$ and $\alpha=1$}: Taking into account the dispersion relation (\ref{Armentano}), we have

\begin{eqnarray}
\partial_{t_{1}}u_{1}^{(1)}-\alpha_{1}\partial_{n_{1}}u_{1}^{(1)}=0,\ \ \ \ \ \alpha_{1}\doteq\frac{\sin\left(\kappa h\right)} {h^2\omega},\label{GruppodiUr}
\end{eqnarray}

\noindent which tell us that $u_{1}^{(1)}$ has the form

\begin{eqnarray}
u_{1}^{(1)}=g\left(\xi,t_{j},j\geq 2\right),\ \ \ \xi\doteq hn_{1}+\frac{\sin\left(\kappa h\right)} {h\omega}t_{1},\label{Parise}
\end{eqnarray}

\noindent where $g$ is an arbitrary function of its arguments going to zero as $\xi\rightarrow\pm\infty$;

\item\emph{Order $\ep^2$ and $\alpha=2$}: Taking into account the dispersion relation (\ref{Armentano}), we have

\begin{eqnarray}
u_{2}^{(2)}=0.\label{GiacomoBoni}
\end{eqnarray}

\item\emph{Order $\ep^3$ and $\alpha=0$}: Taking into account relation (\ref{Pyrotechnia}), we have

\begin{eqnarray}
u_{3}^{(0)}=0;\label{Iuppiter}
\end{eqnarray}

\item\emph{Order $\ep^3$ and $\alpha=1$}: Taking into account the dispersion relation (\ref{Armentano}) and the eqs. (\ref{Pyrotechnia}, \ref{GruppodiUr}), we have

\begin{subequations}
\begin{eqnarray}
\partial_{t_{1}}u_{2}^{(1)}-\alpha_{1}\partial_{n_{1}}u_{2}^{(1)}=-\partial_{t_{2}}u_{1}^{(1)}-\ri\rho_{1}\partial_{\xi}^{2}u_{1}^{(1)}-\ri\rho_{2}|u_{1}^{(1)}|^2,\ \ \ \ \ \ \ \ \ \ \ \ \ \ \ \label{Amor}\\
\rho_{1}\doteq\frac{3+\left(h^2-4\right)\cos\left(\kappa h\right)+\cos\left(2\kappa h\right)} {2h^2\omega\left[h^2-4+4\cos\left(\kappa h\right)\right]},\label{Orma}\ \ \ \ \ \ \ \ \ \ \ \ \ \ \ \ \ \ \ \ \ \ \ \ \ \ \ \ \ \ \ \ \ \ \ \ \\
\rho_{2}\doteq-\frac{4\mu_{2}+\left(5\mu_{1}+3\mu_{3}\right)\cos\left(\kappa h\right)+2\mu_{2}\cos\left(2\kappa h\right)+\mu_{1}\cos\left(3\kappa h\right)} {20\omega},\label{Roma1}\ \ \,
\end{eqnarray}

\begin{eqnarray}
\rho_{2}\doteq-\frac{\left(5\mu_{2}+8\mu_{3}\right)\mu_{1}-\mu_{2}\mu_{3}+\left[\left(11\mu_{1}-5\mu_{2}\right)\mu_{1}+3\left(\mu_{2}+\mu_{3}\right)\mu_{3}\right]\cos\left(\kappa h\right)} {\omega}-\label{Roma2}\ \ \ \ \ \ \ \\
\nonumber-\frac{\left(\mu_{1}\mu_{2}+4\mu_{1}\mu_{3}-2\mu_{2}\mu_{3}\right)\cos\left(2\kappa h\right)+\mu_{1}\left(\mu_{1}-\mu_{2}\right)\cos\left(3\kappa h\right)} {\omega},\ \ \ \ \ \ \ \ \ \ \ \ \ \ \ \ \ \ \ \ \ \\
\rho_{2}\doteq-\frac{3\left(4+h^2\right)\mu_{1}+2\mu_{2}+3\left(1-2\mu_{2}\right)\cos\left(\kappa h\right)+\left[3\left(4+h^2\right)\mu_{1}+4\mu_{2}\right]\cos\left(2\kappa h\right)} {4\omega},\label{Roma3}
\end{eqnarray}
\end{subequations}
where \rref{Roma1} is obtained in the case of the model $Q_1$, \rref{Roma2} for $Q_2$ and \rref{Roma3} for $Q_3$.
 As a consequence of eq. (\ref{GruppodiUr}), the right hand side of eq. (\ref{Amor}) is secular. Hence we have to require that

\begin{subequations}
\begin{eqnarray}
\partial_{t_{1}}u_{2}^{(1)}-\alpha_{1}\partial_{n_{1}}u_{2}^{(1)}=0,\label{Fauno}\ \ \ \ \ \ \ \ \ \ \ \ \ \ \ \ \ \ \ \ \ \ \ \ \ \ \ \,\\
\partial_{t_{2}}u_{1}^{(1)}=-\ri\rho_{1}\partial_{\xi}^{2}u_{1}^{(1)}-\ri\rho_{2}|u_{1}^{(1)}|^2\doteq K_{2}\left[u_{1}^{(1)}\right].\label{Pico} 
\end{eqnarray}
\end{subequations}

\noindent Eq. (\ref{Fauno}) tells us that $u_{2}^{(1)}$ also depends on $\xi$ while  (\ref{Pico}) is an integrable $NLSE$, as from the definitions (\ref{Roma1}, \ref{Roma2}, \ref{Roma3}) we can see that $\rho_{2}$ is a real coefficient in all of the three cases. Hence all the models are $A_{1}-$integrable. 

If we require that our models be  $A_{1}-$linearisable \cite{Burgers}, then we need $\rho_{2}=0$ for any $\kappa$. In this case we have respectively

\begin{subequations}
\begin{align}
&Q_1: \qquad \mu_{1}=\mu_{2}=\mu_{3}=0,\\
&Q_2: \qquad \mu_{1}=\mu_{3}=0,\\
&Q_3: \qquad \mbox{no solution},
\end{align}
\end{subequations}
i.e.   in the first two models only the trivial linear cases are selected while the third model doesn't admit an $A_{1}-$linearisable reduction.

\item\emph{Order $\ep^3$ and $\alpha=2$}: Taking into account the dispersion relation (\ref{Armentano}) and (\ref{Pyrotechnia}, \ref{GiacomoBoni}), we have

\begin{eqnarray}
 u_{3}^{(2)}=0;\label{Luce}
\end{eqnarray}

\item\emph{Order $\ep^3$ and $\alpha=3$}: Taking into account the dispersion relation (\ref{Armentano}), we obtain

\begin{subequations}\label{Saturnia}
\begin{eqnarray}
u_{3}^{(3)}=\alpha_{2}u_{1}^{(1)3},\ \ \ \ \ \ \ \ \ \ \ \ \ \ \ \ \ \ \ \ \ \ \ \ \ \ \ \ \ \ \ \ \ \ \ \ \ \ \ \ \ \ \ \ \ \ \ \ \ \ \ \ \ \ \ \ \ \ \ \ \ \ \ \ \ \ \ \ \ \ \ \ \ \ \ \,\label{Saturnia4}\\
\alpha_{2}\doteq\frac{h^2\left[\mu_{3}+2\mu_{2}\cos\left(\kappa h\right)+2\mu_{1}\cos\left(2\kappa h\right)\right]\cos\left(\kappa h\right)} {20\left[8-2h^2-9\cos\left(\kappa h\right)+\cos\left(3\kappa h\right)\right]},\label{Saturnia1}\ \ \ \ \ \ \ \ \ \ \ \ \ \ \ \ \ \ \ \ \ \ \ \ \ \ \ \\
\alpha_{2}\doteq\frac{h^2\left[\mu_{3}+2\mu_{1}\cos\left(\kappa h\right)\right]\left[\mu_{1}-\mu_{2}+\left(\mu_{2}+\mu_{3}\right)\cos\left(\kappa h\right)+\mu_{1}\cos\left(2\kappa h\right)\right]} {8-2h^2-9\cos\left(\kappa h\right)+\cos\left(3\kappa h\right)},\,\label{Saturnia2}\\
\alpha_{2}\doteq\frac{h^2\left[\left(4+h^2\right)\mu_{1}+2\mu_{2}+\left(1-2\mu_{2}\right)\cos\left(\kappa h\right)+\left(4+h^2\right)\mu_{1}\cos\left(2\kappa h\right)\right]} {4\left[8-2h^2-9\cos\left(\kappa h\right)+\cos\left(3\kappa h\right)\right]},\label{Saturnia3}
\end{eqnarray}
\end{subequations}
where \rref{Saturnia1} is obtained in the case of the model $Q_1$, \rref{Saturnia2} for $Q_2$ and \rref{Saturnia3} for $Q_3$.
\item\emph{Order $\ep^4$ and $\alpha=0$}: Taking into account  (\ref{Pyrotechnia}, \ref{Platone}, \ref{GiacomoBoni}), we get

\begin{eqnarray}
u_{4}^{(0)}=0;\label{PietrodeAngelis}      
\end{eqnarray}

\item\emph{Order $\ep^4$ and $\alpha=1$}: Taking into account the dispersion relation (\ref{Armentano}) and eqs. (\ref{Pyrotechnia}, \ref{GruppodiUr}, \ref{Fauno}, \ref{Pico}), we get

\begin{subequations}
\begin{align}
&\partial_{t_{1}}u_{3}^{(1)}-\alpha_{1}\partial_{n_{1}}u_{3}^{(1)}=-\left(\partial_{t_{3}}u_{1}^{(1)}-K_{3}\left[u_{1}^{(1)}\right]\right)-\left(\partial_{t_{2}}u_{2}^{(1)}-K_{2}^{\prime}\left[u_{1}^{(1)}\right]u_{2}^{(1)}-f_{2}\left(2\right)\right),\label{Talao}\\
&B\doteq\frac{\left[6-8h^2+h^4+2\left(h^2-4\right)\cos\left(\kappa h\right)+2\cos\left(2\kappa h\right)\right]\sin\left(\kappa h\right)} {6h^2\omega\left[h^2-4+4\cos\left(\kappa h\right)\right]^2},\\ \nonumber
&K_{2}^{\prime}\left[u_{1}^{(1)}\right]u_{2}^{(1)}\doteq-\ri\rho_{1}\left[\partial_{\xi}^2u_{2}^{(1)}+\frac{\rho_{2}} {\rho_{1}}\left(u_{1}^{(1)2}\bar u_{2}^{(1)}+2|u_{1}^{(1)}|^2u_{2}^{(1)}\right)\right].\end{align}
\end{subequations}

\noindent In the above relations $K_{3}\left[u_{1}^{(1)}\right]$ \rref{Rutuli2} is the $cmKdV$ flux, the second flux of the $NLSE$ hierarchy, $K_{2}^{\prime}\left[u_{1}^{(1)}\right]u_{2}^{(1)}$ is the Frechet derivative along the direction $u_{2}^{(1)}$ of the $NLSE$  flux $K_{2}\left[u_{1}^{(1)}\right]$ defined by relation (\ref{Pico}) and $f_{2}\left(2\right)$ \rref{Abruzzo1} is a nonlinear forcing term depending on $u_{1}^{(1)}$ and defined by the coefficients $a$ and $b$. As a consequence of  (\ref{GruppodiUr}, \ref{Fauno}) the right hand side of  (\ref{Talao}) is secular, so that
\begin{subequations}
\begin{align}
&\partial_{t_{1}}u_{3}^{(1)}-\alpha_{1}\partial_{n_{1}}u_{3}^{(1)}=0,\label{Mongenet},\\
&\partial_{t_{2}}u_{2}^{(1)}-K_{2}^{\prime}\left[u_{1}^{(1)}\right]u_{2}^{(1)}=-\left(\partial_{t_{3}}u_{1}^{(1)}-K_{3}\left[u_{1}^{(1)}\right]\right)+f_{2}\left(2\right).\label{Ribulsi}
\end{align}
\end{subequations}
Eq. \rref{Mongenet} tells us that $u_{3}^{(1)}$ depends on $\xi$ too while in \rref{Ribulsi}, as a consequence of  (\ref{Pico}), the first term of the right hand side is secular, so that
\begin{subequations}
\begin{align}
&\partial_{t_{2}}u_{2}^{(1)}-K_{2}^{\prime}\left[u_{1}^{(1)}\right]u_{2}^{(1)}=f_{2}\left(2\right),\label{Acca}\\
&\partial_{t_{3}}u_{1}^{(1)}-K_{3}\left[u_{1}^{(1)}\right]=0.\label{Geronta} 
\end{align}
\end{subequations}
\noindent The coefficients $a$ and $b$ of the forcing term $f_{2}  \left(2\right)$ \rref{Abruzzo1} are given by
{\tiny
\begin{subequations}
\begin{align}
a=&\frac{\left[-36\left(h^2-4\right)\mu_{1}+3\left(4-8h^2+h^4\right)\mu_{2}-5\left(h^2-4\right)\mu_{3}\right]\sin\left(\kappa h\right)} {\Delta_{a}}+\label{pippoa1}\\
\nonumber+&\frac{\left[\left(-153+32h^2-4h^4\right)\mu_{1}-6\left(h^2-4\right)\mu_{2}-\left(2+8h^2-h^4\right)\mu_{3}\right]\sin\left(2\kappa h\right)} {\Delta_{a}}-\\
\nonumber-&\frac{\left[23\left(h^2-4\right)\mu_{1}+\left(36-8h^2+h^4\right)\mu_{2}-3\left(h^2-4\right)\mu_{3}\right]\sin\left(3\kappa h\right)} {\Delta_{a}}+\\
\nonumber+&\frac{\left[\left(-42+8h^2-h^4\right)\mu_{1}-3\left(h^2-4\right)\mu_{2}+5\mu_{3}\right]\sin\left(4\kappa h\right)} {\Delta_{a}}-\\
\nonumber-&\frac{3\left(h^2-4\right)\mu_{1}\sin\left(5\kappa h\right)+\mu_{1}\sin\left(6\kappa h\right)} {\Delta_{a}},\\
\nonumber\Delta_{a}=&40\omega\left[h^2-4+4\cos\left(\kappa h\right)\right]\left[3+\left(h^2-4\right)\cos\left(\kappa h\right)+\cos\left(2\kappa h\right)\right],\\
b=&\frac{\left\{12\mu_{2}+\left(h^2-4\right)\mu_{3}+2\left[7\mu_{1}+\left(h^2-4\right)\mu_{2}+5\mu_{3}\right]\cos\left(\kappa h\right)\right.} {\Delta_{b}}+\label{pippob1}\\
\nonumber+&\frac{\left.2\left[\left(h^2-4\right)\mu_{1}+4\mu_{2}\right]\cos\left(2\kappa h\right)+6\mu_{1}\cos\left(3\kappa h\right)\right\}\sin\left(\kappa h\right)} {\Delta_{b}},\\
\nonumber\Delta_{b}=&20\omega\left[h^2-4+4\cos\left(\kappa h\right)\right],\\
a=&\frac{\left\{\left[\left(50-48h^2+7h^4\right)\mu_{2}-46\left(h^2-4\right)\mu_{1}+6\left(4-8h^2+h^4\right)\mu_{3}\right]\mu_{1}+\right.}  {\Delta_{a}}\label{pippoa2}\\
\nonumber&\frac{+\left.\left[\left(70-21h^2+2h^4\right)\mu_{2}-5\left(h^2-4\right)\mu_{3}\right]\mu_{3}\right\}\sin\left(\kappa h\right)} {\Delta_{a}}+\\
\nonumber+&\frac{\left\{\left[\left(1+10h^2\right)\mu_{2}+\left(-157+16h^2-2h^4\right)\mu_{1}\right]\mu_{1}+\right.}  {\Delta_{a}}\\
\nonumber&\frac{+\left.\left[\left(-58+6h^2+h^4\right)\mu_{2}-12\left(h^2-4\right)\mu_{1}+\left(-2-8h^2+h^4\right)\mu_{3}\right]\mu_{3}\right\}\sin\left(2\kappa h\right)} {\Delta_{a}}+\\
\nonumber+&\frac{\left\{-\left[\left(59-19h^2+h^4\right)\mu_{2}+17\left(h^2-4\right)\mu_{1}+2\left(36-8h^2+h^4\right)\mu_{3}\right]\mu_{1}+\right.}  {\Delta_{a}}\\
\nonumber&\frac{\left.+3\left[\left(h^2+4\right)\mu_{2}+\left(h^2-4\right)\mu_{3}\right]\mu_{3}\right\} \sin\left(3\kappa h\right)} {\Delta_{a}}+\\
\nonumber+&\frac{\left\{\left[\left(46-11h^2+h^4\right)\mu_{2}-\left(32-8h^2+h^4\right)\mu_{1}-6\left(h^2-4\right)\mu_{3}\right]\mu_{1}+5\left(\mu_{2}+\mu_{3}\right)\mu_{3}\right\}\sin\left(4\kappa h\right)}  {\Delta_{a}}-\\
\nonumber-&\frac{\left\{\left[\left(13-3h^2\right)\mu_{2}+3\left(h^2-4\right)\mu_{1}\right]\mu_{1}+2\mu_{2}\mu_{3}\right\}\sin\left(5\kappa h\right)+\left(\mu_{1}-\mu_{2}\right)\mu_{1}\sin\left(6\kappa h\right)}  {\Delta_{a}},\\
\nonumber\Delta_{a}=&2\omega\left[h^2-4+4\cos\left(\kappa h\right)\right]\left[3+\left(h^2-4\right)\cos\left(\kappa h\right)+\cos\left(2\kappa h\right)\right],\\
b=&\frac{\left\{\left[\left(h^2-4\right)\mu_{1}-3\left(h^2-7\right)\mu_{2}+16\mu_{3}\right]\mu_{1}+\left[\left(h^2-8\right)\mu_{2}+\left(h^2-4\right)\mu_{3}\right]\mu_{3}\right\}\sin\left(\kappa h\right)} {\Delta_{b}}+\label{pippob2}\\
\nonumber+&\frac{\left\{2\left(7\mu_{1}-6\mu_{2}\right)\mu_{1}+\left[2\left(h^2-4\right)\mu_{1}+\left(13-2h^2\right)\mu_{2}+5\mu_{3}\right]\mu_{3}\right\}\sin\left(2\kappa h\right)} {\Delta_{b}}+\\
\nonumber+&\frac{\left\{\left[\left(h^2-4\right)\mu_{1}-\left(h^2-5\right)\mu_{2}\right]\mu_{1}+2\left(4\mu_{1}-3\mu_{2}\right)\mu_{3}\right\}\sin\left(3\kappa h\right)+3\left(\mu_{1}-\mu_{2}\right)\mu_{1}\sin\left(4\kappa h\right)} {\Delta_{b}},\\
\nonumber\Delta_{b}=&\omega\left[h^2-4+4\cos\left(\kappa h\right)\right],
\end{align}
\begin{align}
a=&-\frac{\left\{4\left(-5+58\mu_{1}+35\mu_{2}\right)+h^2\left[5+\left(26-4h^2+h^4\right)\mu_{1}+2\left(2h^2-21\right)\mu_{2}\right]\right\}\sin\left(\kappa h\right)} {\Delta_{a}}+\label{pippoa3}\\
\nonumber+&\frac{\left[-2+352\mu_{1}+h^4\left(1-22\mu_{1}-2\mu_{2}\right)+116\mu_{2}-4h^2\left(2+3\mu_{2}\right)\right]\sin\left(2\kappa h\right)} {\Delta_{a}}-\\
\nonumber-&\frac{\left\{12\left(1+19\mu_{1}+2\mu_{2}\right)+h^2\left[-3+\left(25-4h^2+h^4\right)\mu_{1}+6\mu_{2}\right]\right\}\sin\left(3\kappa h\right)} {\Delta_{a}}+\\
\nonumber+&\frac{\left[5-3\left(h^4-16\right)\mu_{1}-10\mu_{2}\right]\sin\left(4\kappa h\right)+\left[\left(h^2+4\right)\mu_{1}+4\mu_{2}\right]\sin\left(5\kappa h\right)} {\Delta_{a}},\\
\nonumber\Delta_{a}=&8\omega\left[h^2-4+4\cos\left(\kappa h\right)\right]\left[3+\left(h^2-4\right)\cos\left(\kappa h\right)+\cos\left(2\kappa h\right)\right],\\
b=&\frac{\left[h^2\left(1+7\mu_{1}-2\mu_{2}\right)+4\left(-1+7\mu_{1}+4\mu_{2}\right)\right]\sin\left(\kappa h\right)} {\Delta_{b}}+\label{pippob3}\\
\nonumber+&\frac{\left[5+2\left(h^4-16\right)\mu_{1}+2\left(2h^2-13\right)\mu_{2}\right]\sin\left(2\kappa h\right)+\left[7\left(h^2+4\right)\mu_{1}+12\mu_{2}\right]\sin\left(3\kappa h\right)} {\Delta_{b}},\\
\nonumber\Delta_{b}=&4\omega\left[h^2-4+4\cos\left(\kappa h\right)\right],
\end{align}
\end{subequations}}
where (\ref{pippoa1}, \ref{pippob1}) is obtained in the case of the model $Q_1$, (\ref{pippoa2}, \ref{pippob2}) for $Q_2$ and (\ref{pippoa3}, \ref{pippob3}) for $Q_3$.
 As one can see, in all the three cases the coefficients $a$ and $b$ are real. As a consequence all the three models are $A_{2}-$integrable;
\item\emph{Order $\ep^4$ and $\alpha=2$}: Taking into account the dispersion relation (\ref{Armentano}) and  (\ref{Pyrotechnia}, \ref{Platone}, \ref{GiacomoBoni}, \ref{Luce}), it results
\begin{eqnarray}
u_{4}^{(2)}=0;
\end{eqnarray}
\item\emph{Order $\ep^4$ and $\alpha=3$}: Taking into account the dispersion relation (\ref{Armentano}) and  (\ref{Pyrotechnia}, \ref{GruppodiUr}, \ref{Saturnia}), it results
\begin{eqnarray}
u_{4}^{(3)}=\left(\alpha_{3}u_{2}^{(1)}+\alpha_{4}\partial_{\xi}u_{1}^{(1)}\right)u_{1}^{(1)2};
\end{eqnarray}
As $u_4^{(3)}$ do not enter into the final result, the coefficients $\alpha_3$ and $\alpha_4$ are not explicitly written down here.
\item\emph{Order $\ep^4$ and $\alpha=4$}: Taking into account the dispersion relation (\ref{Armentano}) and (\ref{GiacomoBoni}), it results
\begin{eqnarray}
u_{4}^{(4)}=0.
\end{eqnarray}
\item\emph{Order $\ep^4$ and $\alpha=0$}: Taking into account (\ref{Pyrotechnia}, \ref{Platone}, \ref{GiacomoBoni}, \ref{Iuppiter}, \ref{Luce}), we get
\begin{eqnarray}
u_{5}^{(0)}=0;\label{Incognita}      
\end{eqnarray}
\item\emph{Order $\ep^5$ and $\alpha=1$}: Taking into account the dispersion relation (\ref{Armentano}) and (\ref{Pyrotechnia}, \ref{Platone}, \ref{GruppodiUr}, \ref{GiacomoBoni}, \ref{Fauno}, \ref{Pico}, \ref{Saturnia}, \ref{Mongenet}, \ref{Acca}, \ref{Geronta}), we get
\begin{subequations}
\begin{align}
&\partial_{t_{1}}u_{4}^{(1)}-\alpha_{1}\partial_{n_{1}}u_{4}^{(1)}=-\left(\partial_{t_{2}}u_{3}^{(1)}-K_{2}^{\prime}\left[u_{1}^{(1)}\right]u_{3}^{(1)}-f_{2}\left(3\right)\right)-\label{Pallanteo}\\
\nonumber&-\left(\partial_{t_{3}}u_{2}^{(1)}-K_{3}^{\prime}\left[u_{1}^{(1)}\right]u_{2}^{(1)}-f_{3}\left(2\right)\right)-\left(\partial_{t_{4}}u_{1}^{(1)}-K_{4}\left[u_{1}^{(1)}\right]\right),\\
&C\doteq\frac{7\left(5-8h^2+h^4\right)+\left(h^2-4\right)\left(14-8h^2+h^4\right)\cos\left(\kappa h\right)} {24h^2\omega\left[h^2-4+4\cos\left(\kappa h\right)\right]^3}+\\
&\nonumber\ \ \,+\frac{\left(28+8h^2-h^4\right)\cos\left(2\kappa h\right)+2\left(h^2-4\right)\cos\left(3\kappa h\right)+\cos\left(4\kappa h\right)} {24h^2\omega\left[h^2-4+4\cos\left(\kappa h\right)\right]^3}.
\end{align}
\end{subequations}
\noindent In the above relations $K_{4}\left[u_{1}^{(1)}\right]$ \rref{Rutuli3} is the is the third flux of the $NLSE$ hierarchy, $K_{3}^{\prime}\left[u_{1}^{(1)}\right]u_{2}^{(1)}$ is the Frechet derivative along the direction $u_{2}^{(1)}$ of the $cmKdV$  flux $K_{3}\left[u_{1}^{(1)}\right]$ defined by relation (\ref{Rutuli2}) and $f_{2}\left(3\right)$, $f_{3}\left(2\right)$ are the nonlinear forcing terms defined in (\ref{Abruzzo2}, \ref{Abruzzo1}). As a consequence of  (\ref{GruppodiUr}, \ref{Fauno}, \ref{Mongenet}) the right hand side of (\ref{Pallanteo}) is secular, so that
\begin{subequations}
\begin{align}
&\partial_{t_{1}}u_{4}^{(1)}-\alpha_{1}\partial_{n_{1}}u_{4}^{(1)}=0,\\
&\partial_{t_{2}}u_{3}^{(1)}-K_{2}^{\prime}\left[u_{1}^{(1)}\right]u_{3}^{(1)}=-\left(\partial_{t_{3}}u_{2}^{(1)}-K_{3}^{\prime}\left[u_{1}^{(1)}\right]u_{2}^{(1)}-f_{3}\left(2\right)\right)-\label{Evandro}\\
\nonumber&\ \ \ \ \ \ \ \ \ \ \ \ \ \ \ \ \ \ \ \ \ \ \ \ \ \ \ \ \ \ \ \ \ -\left(\partial_{t_{4}}u_{1}^{(1)}-K_{4}\left[u_{1}^{(1)}\right]\right)+f_{2}\left(3\right).
\end{align}
\end{subequations}
\noindent The first relation tells us that $u_{4}^{(1)}$ depends on $\xi$ too while in (\ref{Evandro}), as a consequence of  (\ref{Acca}) and of
\begin{eqnarray}
\nonumber\left(\partial_{t_{2}}-K_{2}^{\prime}\left[u_{1}^{(1)}\right]\right)f_{3}\left(2\right)=\left(\partial_{t_{3}}-K_{3}^{\prime}\left[u_{1}^{(1)}\right]\right)f_{2}\left(2\right),
\end{eqnarray}
\noindent the first term on the right hand side is secular. Moreover also the second term on the right hand side is secular because, when we equal it to zero, we obtain a generalized symmetry the $NLSE$,  the third equation of the corresponding hierarchy. Hence
\begin{subequations}
\begin{align}
&\partial_{t_{2}}u_{3}^{(1)}-K_{2}^{\prime}\left[u_{1}^{(1)}\right]u_{3}^{(1)}=f_{2}\left(3\right),\\
&\partial_{t_{3}}u_{2}^{(1)}-K_{3}^{\prime}\left[u_{1}^{(1)}\right]u_{2}^{(1)}=-\left(\partial_{t_{4}}u_{1}^{(1)}-K_{4}\left[u_{1}^{(1)}\right]\right)+f_{3}\left(2\right).\label{Maro}
\end{align}
\end{subequations}
\noindent Finally the first term on the right hand side of  (\ref{Maro}) is secular because, when we equal it to zero, we obtain a generalized symmetry the $cmKdV$ equation as both equations belong to the same $NLSE$ hierarchy. Hence
\begin{subequations}
\begin{align}
&\partial_{t_{3}}u_{2}^{(1)}-K_{3}^{\prime}\left[u_{1}^{(1)}\right]u_{2}^{(1)}=f_{3}\left(2\right),\\
&\partial_{t_{4}}u_{1}^{(1)}-K_{4}\left[u_{1}^{(1)}\right]=0.
\end{align}
\end{subequations}
\noindent The real and imaginary parts ($R_{j}$, $I_{j}$) of the coefficients $\tau_{j}$, $j=1$,\ldots,12 of the forcing term $f_{2}(3)$ are given by
\begin{align}
&R_{1}=R_{2}=R_{3}=R_{4}=R_{5}=R_{8}=R_{9}=0,\ \ \ \ R_{10}=R_{12},\\
&I_{6}=I_{7}=I_{10}=I_{11}=I_{12}=0,\ \ \ \ I_{9}=2I_{8},
\end{align}
\noindent independently from the model.  
{\tiny
\begin{subequations} \label{last1}
\begin{align}
I_{1}=&\frac{2\alpha_{2}\mu_{2}\rho_{1}^{2}-5\left(2\rho_{1}^2\rho_{2}+3aB\omega-6C\rho_{2}\omega\right)\rho_{2}+\alpha_{2}\rho_{1}^{2}\left\{\mu_{2}\left[3\cos\left(2\kappa h\right)+\cos\left(4\kappa h\right)\right]+\right.} {20\rho_{1}^2\omega}\\
\nonumber&\frac{\left.+\left[\mu_{3}+2\mu_{1}\cos\left(2\kappa h\right)\right]\left[2\cos\left(\kappa h\right)+\cos\left(3\kappa h\right)\right]\right\}} {20\rho_{1}^2\omega},\\
I_{2}=&\frac{2\mu_{2}\rho_{1}+20\left[\left(a+2b\right)\alpha_{1}\rho_{1}+\left(3B\alpha_{1}-2\rho_{1}^{2}+2C\omega\right)\rho_{2}-3bB\omega\right]+} {20\rho_{1}\omega}\\
\nonumber&\frac{+6\mu_{1}\cos\left(\kappa h\right)-\mu_{2}\cos\left(2\kappa h\right)-2\mu_{1}\cos\left(3\kappa h\right)} {20\omega},\\
I_{3}=&\frac{2a\alpha_{1}\rho_{1}+2\left(3B\alpha_{1}-2\rho_{1}^{2}+4C\omega\right)\rho_{2}-3aB\omega+} {2\rho_{1}\omega}\\
\nonumber&\frac{+\left[4\mu_{1}+\mu_{3}+3\mu_{2}\cos\left(\kappa h\right)+2\mu_{1}\cos\left(2\kappa h\right)\right]\cos\left(\kappa h\right)} {20\omega},\\
I_{4}=&\frac{2\left[\left(20b\alpha_{1}+\mu_{2}\right)\rho_{1}+20C\rho_{2}\omega\right]+\left[\left(5\mu_{1}+\mu_{3}\right)\cos\left(\kappa h\right)+\mu_{2}\cos\left(2\kappa h\right)+\mu_{1}\cos\left(3\kappa h\right)\right]\rho_{1}} {40\rho_{1}\omega},\\
I_{5}=&\frac{20\left(2\alpha_{1}\rho_{1}-3B\omega\right)a-\mu_{2}\rho_{1}+40\left(3B\alpha_{1}-\rho_{1}^2+3C\omega\right)\rho_{2}+2\left[\mu_{2}+2\mu_{1}\cos\left(\kappa h\right)\right]\rho_{1}\cos\left(2\kappa h\right)} {40\rho_{1}\omega},\\
R_{6}=&\frac{\left(2\alpha_{1}\rho_{1}-3B\omega\right)\rho_{2}} {\rho_{1}\omega}-\frac{\left[4\mu_{1}+\mu_{3}+3\mu_{2}\cos\left(\kappa h\right)+2\mu_{1}\cos\left(2\kappa h\right)\right]\sin\left(\kappa h\right)} {10\omega},\\
R_{7}=&\frac{20\alpha_{1}\rho_{2}+\left[\mu_{3}+2\mu_{1}\cos\left(2\kappa h\right)\right]\sin\left(\kappa h\right)+\mu_{2}\sin\left(2\kappa h\right)} {20\omega},\\
I_{8}=&\frac{\left(5\mu_{1}+3\mu_{3}\right)\cos\left(\kappa h\right)+2\left[2+\cos\left(2\kappa h\right)\right]\mu_{2}+\mu_{1}\cos\left(3\kappa h\right)} {20\omega},\\
R_{10}=&\frac{\left(2\alpha_{1}\rho_{1}-3B\omega\right)\rho_{2}} {\rho_{1}\omega}-\frac{\left[4\mu_{1}+\mu_{3}+3\mu_{2}\cos\left(\kappa h\right)+2\mu_{1}\cos\left(2\kappa h\right)\right]\sin\left(\kappa h\right)} {10\omega},\\
R_{11}=&\frac{20\alpha_{1}\rho_{2}+\left[\mu_{3}+2\mu_{1}\cos\left(2\kappa h\right)\right]\sin\left(\kappa h\right)+\mu_{2}\sin\left(2\kappa h\right)} {10\omega},\end{align}
\end{subequations}}
for the model $Q_1$
{\tiny
\begin{subequations} \label{last2}
\begin{align}
I_{1}=&\frac{4\left(3\mu_{1}\mu_{2}+4\mu_{1}\mu_{3}-\mu_{2}\mu_{3}\right)\alpha_{2}\rho_{1}^{2}-\left(2\rho_{1}^{2}\rho_{2}+3aB\omega-6C\rho_{2}\omega\right)\rho_{2}+} {4\rho_{1}^2\omega}\\
\nonumber&\frac{+4\alpha_{2}\rho_{1}^{2}\left\{\left[\left(7\mu_{1}-2\mu_{2}\right)\mu_{1}+2\left(\mu_{2}+\mu_{3}\right)\mu_{3}\right]\cos\left(\kappa h\right)+3\mu_{1}\left(\mu_{2}+2\mu_{3}\right)\cos\left(2\kappa h\right)+\right.} {4\rho_{1}^2\omega}\\
\nonumber&\frac{\left.+\left[\left(4\mu_{1}-3\mu_{2}\right)\mu_{1}+\left(\mu_{2}+\mu_{3}\right)\mu_{3}\right]\cos\left(3\kappa h\right)+\left(\mu_{1}-\mu_{2}\right)\left[2\mu_{3}\cos\left(4\kappa h\right)+\mu_{1}\cos\left(5\kappa h\right)\right]\right\}} {4\rho_{1}^2\omega},\\
I_{2}=&\frac{\left(a+2b\right)\alpha_{1}\rho_{1}+\left(3B\alpha_{1}-2\rho_{1}^{2}+2C\omega\right)\rho_{2}-3bB\omega+2\left(\mu_{2}+2\mu_{3}\right)\mu_{1}\rho_{1}+} {\rho_{1}\omega}\\
\nonumber&\frac{+2\left\{3\mu_{1}^{2}\cos\left(\kappa h\right)+\left(\mu_{2}-\mu_{1}\right)\left[\mu_{3}\cos\left(2\kappa h\right)+\mu_{1}\cos\left(3\kappa h\right)\right]\right\}} {\omega},\\
I_{3}=&\frac{2a\alpha_{1}\rho_{1}+2\left(3B\alpha_{1}-2\rho_{1}^{2}+4C\omega\right)\rho_{2}-3aB\omega+} {2\rho_{1}\omega}\\
\nonumber&\frac{+\left\{2\left(3\mu_{1}-\mu_{2}\right)\mu_{1}+\left(\mu_{2}+\mu_{3}\right)\mu_{3}+2\left[\left(\mu_{2}+3\mu_{3}\right)\mu_{1}-\mu_{2}\mu_{3}\right]\cos\left(\kappa h\right)+2\left(\mu_{1}-\mu_{2}\right)\mu_{1}\cos\left(2\kappa h\right)\right\}\cos\left(\kappa h\right)} {\omega},\\
I_{4}=&\frac{2\left[b\alpha_{1}+\left(\mu_{2}+2\mu_{3}\right)\mu_{1}\right]\rho_{1}+2C\rho_{2}\omega+\left\{\left[\left(7\mu_{1}-3\mu_{2}\right)\mu_{1}+\left(\mu_{2}+\mu_{3}\right)\mu_{3}\right]\cos\left(\kappa h\right)+\right.} {2\rho_{1}\omega}\\
\nonumber&\frac{\left.+\left(\mu_{1}-\mu_{2}\right)\left[2\mu_{3}\cos\left(2\kappa h\right)+\mu_{1}\cos\left(3\kappa h\right)\right]\right\}\rho_{1}} {2\rho_{1}\omega},\\
I_{5}=&\frac{6\left(B\alpha_{1}+C\omega\right)\rho_{2}-2\left[\left(\mu_{1}-\mu_{2}\right)\mu_{3}+\rho_{1}\rho_{2}\right]\rho_{1}+\left(2\alpha_{1}\rho_{1}-3B\omega\right)a+} {2\rho_{1}\omega}\\
\nonumber&\frac{+2\left[\left(\mu_{1}+2\mu_{2}\right)\cos\left(\kappa h\right)+\left(\mu_{2}+2\mu_{3}\right)\cos\left(2\kappa h\right)+\left(\mu_{1}-\mu_{2}\right)\cos\left(3\kappa h\right)\right]\mu_{1}\rho_{1}} {2\rho_{1}\omega},\\
R_{6}=&\frac{\left(2\alpha_{1}\rho_{1}-3B\omega\right)\rho_{2}-2\left\{2\left(3\mu_{1}-\mu_{2}\right)\mu_{1}+\left(\mu_{2}+\mu_{3}\right)\mu_{3}+\right.} {\rho_{1}\omega}\\
\nonumber&\frac{\left.+2\left[\left(\mu_{2}+3\mu_{3}\right)\mu_{1}-\mu_{2}\mu_{3}\right]\cos\left(\kappa h\right)+2\mu_{1}\left(\mu_{1}-\mu_{2}\right)\cos\left(2\kappa h\right)\right\}\rho_{1}\sin\left(\kappa h\right)} {\rho_{1}\omega},
\end{align}
\end{subequations}}
{\tiny \begin{subequations} \label{last21}
\begin{align}
R_{7}=&\frac{\alpha_{1}\rho_{2}+\left[\left(\mu_{1}-3\mu_{2}\right)\mu_{1}+\left(\mu_{2}+\mu_{3}\right)\mu_{3}\right]\sin\left(\kappa h\right)+\left(\mu_{1}-\mu_{2}\right)\left[2\mu_{3}\sin\left(2\kappa h\right)+\mu_{1}\sin\left(3\kappa h\right)\right]} {\omega},\\
I_{8}=&\frac{\left(5\mu_{2}+8\mu_{3}\right)\mu_{1}-\mu_{2}\mu_{3}+\left[\left(11\mu_{1}-5\mu_{2}\right)\mu_{1}+3\left(\mu_{2}+\mu_{3}\right)\mu_{3}\right]\cos\left(\kappa h\right)+} {\omega}\\
\nonumber&\frac{+\left[\left(\mu_{2}+4\mu_{3}\right)\mu_{1}-2\mu_{2}\mu_{3}\right]\cos\left(2\kappa h\right)+\left(\mu_{1}-\mu_{2}\right)\mu_{1}\cos\left(3\kappa h\right)} {\omega},\\
R_{10}=&\frac{\left(2\alpha_{1}\rho_{1}-3B\omega\right)\rho_{2}-2\left\{2\left(3\mu_{1}-\mu_{2}\right)\mu_{1}+\left(\mu_{2}+\mu_{3}\right)\mu_{3}+\right.} {\rho_{1}\omega}\\
\nonumber&\frac{\left.+2\left[\left(\mu_{2}+3\mu_{3}\right)\mu_{1}-\mu_{2}\mu_{3}\right]\cos\left(\kappa h\right)+2\left(\mu_{1}-\mu_{2}\right)\mu_{1}\cos\left(2\kappa h\right)\right\}\rho_{1}\sin\left(\kappa h\right)} {\rho_{1}\omega},\\
R_{11}=&\frac{2\left\{\alpha_{1}\rho_{2}+\left[\left(\mu_{1}-3\mu_{2}\right)\mu_{1}+\left(\mu_{2}+\mu_{3}\right)\mu_{3}\right]\sin\left(\kappa h\right)+\left(\mu_{1}-\mu_{2}\right)\left[2\mu_{3}\sin\left(2\kappa h\right)+\mu_{1}\sin\left(3\kappa h\right)\right]\right\}} {\omega}.
\end{align}
\end{subequations}}
for the model $Q_2$ and 
{\tiny
\begin{subequations}\label{last3}
\begin{align}
I_{1}=&\frac{\left[\left(h^2+4\right)\mu_{1}+2\mu_{2}\right]\alpha_{2}\rho_{1}^{2}-\left(2\rho_{1}^{2}\rho_{2}+3aB\omega-6C\rho_{2}\omega\right)\rho_{2}+\alpha_{2}\rho_{1}^{2}\left\{2\left(1-2\mu_{2}\right)\cos\left(\kappa h\right)+\right.} {4\rho_{1}^2\omega}\\
\nonumber&\frac{\left.+3\left(h^2+4\right)\mu_{1}\cos\left(2\kappa h\right)+\left(1-2\mu_{2}\right)\cos\left(3\kappa h\right)+\left[\left(h^2+4\right)\mu_{1}+2\mu_{2}\right]\cos\left(4\kappa h\right)\right\}} {4\rho_{1}^2\omega},\\
I_{2}=&\frac{2\left[\left(a+2b\right)\alpha_{1}\rho_{1}+\left(3B\alpha_{1}-2\rho_{1}^{2}+2C\omega\right)\rho_{2}-3bB\omega\right]+} {2\rho_{1}\omega}\\
\nonumber&\frac{+\left(h^2+4\right)\mu_{1}-4\rho_{1}\rho_{2}-\left[\left(h^2+4\right)\mu_{1}+2\mu_{2}\right]\cos\left(2\kappa h\right)} {2\omega},\\
I_{3}=&\frac{2a\alpha_{1}\rho_{1}+2\left(3B\alpha_{1}-2\rho_{1}^{2}+4C\omega\right)\rho_{2}-3aB\omega+} {2\rho_{1}\omega}\\
\nonumber&\frac{+\left\{1-2\mu_{2}+4\left[\left(h^2+4\right)\mu_{1}+\mu_{2}\right]\cos\left(\kappa h\right)\right\}\cos\left(\kappa h\right)} {4\omega},\\
I_{4}=&\frac{2\left[4b\alpha_{1}+\left(h^2+4\right)\mu_{1}\right]\rho_{1}+8C\rho_{2}\omega+\left\{\left(1-2\mu_{2}\right)\cos\left(\kappa h\right)+2\left[\left(h^2+4\right)\mu_{1}+2\mu_{2}\right]\cos\left(2\kappa h\right)\right\}\rho_{1}} {8\rho_{1}\omega}.\\
I_{5}=&\frac{12\left(B\alpha_{1}+C\omega\right)\rho_{2}-\left[\left(h^2+4\right)\mu_{1}+2\left(\mu_{2}+2\rho_{1}\rho_{2}\right)\right]\rho_{1}+2\left(2\alpha_{1}\rho_{1}-3B\omega\right)a+\left(h^2+4\right)\mu_{1}\rho_{1}\cos\left(2\kappa h\right)} {4\rho_{1}\omega},\\
R_{6}=&\frac{\left(2\alpha_{1}\rho_{1}-3B\omega\right)\rho_{2}} {\rho_{1}\omega}+\frac{\left(2\mu_{2}-1\right)\sin\left(\kappa h\right)-2\left[\left(h^2+4\right)\mu_{1}+\mu_{2}\right]\sin\left(2\kappa h\right)} {2\omega},\\
R_{7}=&\frac{4\alpha_{1}\rho_{2}+\left(1-2\mu_{2}\right)\sin\left(\kappa h\right)+2\left[\left(h^2+4\right)\mu_{1}+2\mu_{2}\right]\sin\left(2\kappa h\right)} {4\omega},\\
I_{8}=&\frac{3\left(h^2+4\right)\mu_{1}+2\mu_{2}+3\left(1-2\mu_{2}\right)\cos\left(\kappa h\right)+\left[3\left(h^2+4\right)\mu_{1}+4\mu_{2}\right]\cos\left(2\kappa h\right)} {4\omega},\\
R_{10}=&\frac{\left(2\alpha_{1}\rho_{1}-3B\omega\right)\rho_{2}} {\rho_{1}\omega}-\frac{\left(1-2\mu_{2}\right)\sin\left(\kappa h\right)+2\left[\left(h^2+4\right)\mu_{1}+\mu_{2}\right]\sin\left(2\kappa h\right)} {2\omega},\\
R_{11}=&\frac{4\alpha_{1}\rho_{2}+\sin (\kappa h) [1-2 \mu_2] + 2 \sin(2 \kappa h) [(h^2 + 4) \mu_1 + 2 \mu_2]} {2\omega},
\end{align}
\end{subequations}}
for the model $Q_3$. As the coefficients (\ref{last1}, \ref{last2}, \ref{last21}, \ref{last3}) do not satisfy the algebraic relations \rref{Siculi}, for any $Q$ the $A_3$ integrability is never satisfied. 

\end{itemize}


\end{document}